\documentclass[11pt]{article}
\usepackage{amsmath,amssymb,amsthm,mathtools}
\usepackage{accanthis,newpxmath}
\usepackage{microtype}
\usepackage[hidelinks]{hyperref}
\usepackage{enumitem}

\newtheorem{theorem}{Theorem}[section]
\newtheorem{proposition}[theorem]{Proposition}

\theoremstyle{definition}
\newtheorem{remark}[theorem]{Remark}
\theoremstyle{plain}
\newtheorem{definition}[theorem]{Definition}

\newcommand{\dd}{\mathrm{d}}
\newcommand{\1}{\mathbf{1}}

\title{Loss-versus-rebalancing with external proportional transaction costs}
\author{Masaaki Fukasawa\thanks{The University of Osaka}}
\date{}

\begin{document}
\maketitle

\begin{abstract}
We study liquidity provision in automated market makers when trading in the external market incurs proportional transaction costs. We consider both separately collected fees and fees retained in reserves. The benchmark is a self-financing strategy holding the same total risky inventory and trading at the external bid and ask. Under continuous external prices and pure arbitrage, the pool and benchmark have identical liquidation wealth whenever the pool fee or the external spread is positive. Thus loss-versus-rebalancing relative to this transaction-cost benchmark is zero, even without a pool fee if the external spread is positive. We then construct retained-fee utility-indifference pools that implement admissible decreasing inventory schedules, and geometric-mean pools that implement feasible constant no-trade boundaries for portfolio weights. These results offer an investment interpretation of liquidity provision: by choosing the pool design, a liquidity provider can specify a trading strategy whose rebalancing is carried out by arbitrageurs.

\end{abstract}


\section{Introduction}

Decentralized exchanges provide a flexible platform for liquidity providers to choose how their capital is allocated and traded. Asset pairs, fee levels, and liquidity ranges determine how a position responds to market prices; more general pool designs also allow providers to specify reserve curves and inventory rules. Arbitrageurs link pool prices to external prices and execute the resulting changes in holdings. Liquidity provision can therefore implement an investment strategy through a pricing rule. To assess this implementation, one must compare its wealth with that of a strategy trading directly in the external market.

Two benchmarks are commonly used. Impermanent loss compares a pool with a buy-and-hold position, while loss-versus-rebalancing (LVR) compares it with a self-financing strategy holding the same risky-asset inventory~\cite{LVR}. In a frictionless external market, fee-free liquidity provision incurs a quadratic-variation loss. Fukasawa et al.~\cite{FMWLP} show that a positive pool fee eliminates positive LVR under continuous prices for a general class of constant function market makers, including concentrated-liquidity pools.

We study the corresponding comparison when external trading incurs proportional transaction costs. The benchmark trades at the external bid and ask, and both positions are valued by external liquidation. We treat two fee mechanisms: separate accounts, as in Uniswap v3, and fees retained in reserves, as in Uniswap v2. Separate fees leave the active reserves on a fixed curve. Retained fees change the level of a fixed utility function, with geometric-mean pools~\cite{FMWG3M} as a special case. The terms ``Uniswap v3 type'' and ``Uniswap v2 type'' refer throughout to fee collection, rather than to a particular reserve curve. We exclude protocol fee deductions and liquidity deposits or withdrawals.

For a pool fee $\tau$ and external cost parameter $\lambda$, the no-arbitrage band is governed by $(1-\tau)(1-\lambda)$. Either friction can support continuous finite-variation reserve dynamics. Under pure arbitrage, trades occur at the band boundaries, where the pool's execution price equals the corresponding external execution price. This yields our main result: the pool and its rebalancing benchmark have identical liquidation wealth. In particular, transaction-cost LVR is zero without a pool fee when the external spread is positive. An external hedge nevertheless incurs the spread integrated against inventory turnover. This cost is shared by the pool and benchmark and hence cancels from their relative comparison.

The wealth identity also gives a constructive approach to investment implementation. We design retained-fee utility-indifference pools whose total risky inventory follows a prescribed decreasing function of the internal price. We then use geometric-mean pools to implement feasible constant no-trade boundaries for portfolio weights. Retaining fees makes the actual reserves the entire strategy portfolio, so these implementations require no separate fee-reinvestment rule. The two constructions illustrate how liquidity providers can choose a trading strategy through pool geometry and have arbitrageurs carry out its rebalancing.

\section{Pool and external market}

An automated market maker (AMM) executes trades against a pool of assets supplied by liquidity providers. A trader deposits one asset and receives the other according to a prescribed pricing rule. A typical design is a constant function market maker (CFMM): a trading function assigns a value to each reserve combination, and the swap output is chosen to preserve that value after adjusting the input for the trading fee. The combinations with the same value form a reserve curve. A utility function can serve as the trading function, giving a utility-indifference AMM.

The pricing rule is specified by the trading function and fee, together with how fees are collected. Fees may be held in separate accounts or retained in the active reserves. This choice determines the reserves used to price subsequent trades. Within the available pool design, liquidity providers can therefore shape their investment strategy by choosing the trading function, fee, and fee-collection mechanism.

We study pools containing a risky asset and a numeraire. Arbitrageurs trade between the pool and an external market whenever the price difference covers the pool fee and external transaction costs. After specifying the external market, we describe separately collected fees and then fees retained in reserves. The first mechanism is naturally described by a fixed reserve curve; the second uses a trading function whose level changes as fees accumulate.

\paragraph{External market.}
Let $S^*>0$ be a continuous external reference price. The external bid and ask are
\begin{equation}\label{eq:external-spread}
S^b=(1-\lambda)S^*,\qquad
S^a=\frac{S^*}{1-\lambda},\qquad \lambda\in[0,1).
\end{equation}
The wealth identities below also apply to general continuous external bid and ask processes; the proportional specification is used to construct the reflected dynamics.

\subsection{Separate fees: Uniswap v3 type}\label{sec:separate}

This subsection extends the separate-fee framework of~\cite{FMWLP} to an external market with proportional transaction costs. The reserve-curve pricing rule is unchanged; the external bid and ask determine the no-arbitrage band and the boundary conditions for reserve trading.

\subsubsection{Reserve curve and execution prices}

Let $(X,Y)$ denote the active reserves of the risky asset and numeraire. With fees collected separately, these reserves remain on a fixed reserve curve. Following~\cite{FMWLP}, let $I\subset[0,\infty)$ be an interval and let
\[
f:I\longrightarrow[0,\infty)
\]
be twice continuously differentiable, strictly convex and decreasing on the interior of $I$. We work on the active range, where $X$ lies in this interior, $Y>0$, and $f'(X)<0$. The reserves satisfy
\[
Y=f(X),\qquad S=s(X):=-f'(X).
\]
The function $f$ is the reserve curve and $S$ is the internal price. The pool charges a fraction $\tau\in[0,1)$ of the incoming asset as a fee, held outside the active reserves. The remaining input and the output move the active reserves along the graph of $f$. The ask is the numeraire paid per unit of risky asset bought from the pool, in the limit of a small trade; the bid is the numeraire received per unit sold to it. To buy $z>0$ units, a trader pays $w(z)$ units of numeraire, of which only $(1-\tau)w(z)$ enters the active reserve. Hence
\[
f(X-z)=f(X)+(1-\tau)w(z),
\qquad
A:=\lim_{z\downarrow0}\frac{w(z)}{z}
=\frac{-f'(X)}{1-\tau}=\frac{S}{1-\tau}.
\]
Conversely, selling $z>0$ units to the pool increases its active risky reserve by $(1-\tau)z$. The numeraire payout is $v(z)=f(X)-f(X+(1-\tau)z)$, so
\[
B:=\lim_{z\downarrow0}\frac{v(z)}{z}
=-(1-\tau)f'(X)=(1-\tau)S.
\]
These are marginal execution prices; a finite trade generally has a different average price.

We write $X^f,Y^f$ for cumulative fees and
\[
Q:=X+X^f
\]
for the liquidity provider's total holding of the risky asset.

\begin{remark}[Utility-indifference representation]
A utility-indifference pool is obtained by taking $f$ to describe a level set of a smooth, strictly concave utility $u$ with positive partial derivatives, as in~\cite{FMWLP}. The separate-fee mechanism requires only this fixed curve.
\end{remark}

\begin{remark}[Concentrated liquidity and multiple providers]
The reserve-curve description also covers a pool with multiple providers supplying concentrated liquidity over different price ranges. As shown in~\cite[Section~6]{FMWLP}, with a common fee rate and a common internal price among active positions, their liquidity can be aggregated into that of a representative provider. The curve then describes the pool's aggregate reserves. The smooth description applies between range boundaries; when positions become active or inactive, the corresponding curve segments are joined with the appropriate endpoint conventions.
\end{remark}

\subsubsection{Example: a geometric-mean market maker}\label{ex:gmm-separate}
Fix a weight $\alpha\in(0,1)$ and positive initial reserves $(X_0,Y_0)$. A geometric-mean market maker (GMM) with separate fees keeps its active reserves on the level set
\[
X^\alpha Y^{1-\alpha}=X_0^\alpha Y_0^{1-\alpha}.
\]
Writing $k=\alpha/(1-\alpha)$ and $C=X_0^kY_0$, this gives the concrete reserve curve
\begin{equation}\label{eq:gmm-fixed-curve}
f(x)=Cx^{-k},\qquad x>0.
\end{equation}
Thus $f(x)$ is the numeraire reserve required when the active risky reserve is $x$. Its derivatives satisfy
\[
f'(x)=-kCx^{-k-1}<0,\qquad
f''(x)=k(k+1)Cx^{-k-2}>0,
\]
and its internal price and reserve--price schedule are
\[
S=k\frac{Y}{X},\qquad
X=\left(\frac{kC}{S}\right)^{1/(k+1)}.
\]
The equal-weight case $\alpha=1/2$ gives $k=1$, $XY=C$, and $f(x)=C/x$.

For example, suppose a trader deposits $z>0$ units of the risky asset into a pool with active reserves $(x,y)$ and $y=Cx^{-k}$. Only $(1-\tau)z$ enters the active risky reserve. The new active reserves and the fee increment are
\[
x'=x+(1-\tau)z,\qquad
y'=C(x')^{-k},\qquad \Delta X^f=\tau z.
\]
The trader receives $y-y'$ units of numeraire. The active curve level remains $(x')^ky'=C$, and the fee is held outside that curve.

\subsubsection{No-arbitrage band and reserve dynamics}

An arbitrageur who buys the risky asset from the pool can liquidate it externally at $S^b$. An arbitrageur who sells it to the pool must first buy it externally at $S^a$. The no-arbitrage conditions are therefore
\begin{equation}\label{eq:na}
A\geq S^b,\qquad B\leq S^a.
\end{equation}
Equivalently, with
\[
\gamma:=(1-\tau)(1-\lambda),
\]
they are
\begin{equation}\label{eq:band}
\gamma S^*\leq S\leq \frac{S^*}{\gamma}.
\end{equation}
Thus the reserve dynamics have the same no-arbitrage band as a frictionless external market with effective fee
\[
\tau_{\rm eff}=1-\gamma=\tau+\lambda-\tau\lambda.
\]
This observation concerns the band only: fee income is still determined by $\tau$.

When $\tau+\lambda>0$, the band has positive width. Starting from reserves in the no-arbitrage band, reflection in the corresponding inventory interval gives continuous reserves of finite variation while the construction remains in the active range. The smooth analysis is stopped if a reserve reaches an endpoint; continuing a concentrated-liquidity position requires its endpoint convention.

Under this positive-width condition, continuous pure arbitrage flow is supported on the boundaries in the following sense:
\begin{equation}\label{eq:support}
(A-S^b)\dd X^\downarrow=0,\qquad
(S^a-B)\dd X^\uparrow=0,
\end{equation}
where $X=X_0+X^\uparrow-X^\downarrow$ and the increasing processes start at zero and have mutually singular increment measures. The pool sells only when $A=S^b$ and buys only when $B=S^a$: away from the corresponding boundary, the arbitrage trade would lose money.

\subsubsection{Fee accounts and liquidation wealth}

If the active risky reserve increases by $\dd X^\uparrow$, the trader deposits $\dd X^\uparrow/(1-\tau)$ and the difference enters the fee account. Writing $Y=Y_0+Y^\uparrow-Y^\downarrow$ for its Jordan decomposition, the same bookkeeping applies to incoming numeraire, giving
\begin{equation}\label{eq:fees}
\dd X^f=\frac{\tau}{1-\tau}\dd X^\uparrow,\qquad
\dd Y^f=\frac{\tau}{1-\tau}\dd Y^\uparrow.
\end{equation}
Since $Y=f(X)$ and $X$ has finite variation,
\begin{equation}\label{eq:Y}
\dd Y=-S\dd X=S\dd X^\downarrow-S\dd X^\uparrow.
\end{equation}
Adding the fee balances to the active reserves gives
\begin{align}
\dd Q^\uparrow&=\frac{1}{1-\tau}\dd X^\uparrow,
\qquad \dd Q^\downarrow=\dd X^\downarrow, \label{eq:Q}\\
\dd(Y+Y^f)&=A\dd X^\downarrow-S\dd X^\uparrow. \label{eq:cash}
\end{align}

The first identity records the trader's gross risky-asset deposit; no risky-asset fee is collected when the pool sells. The cash identity includes the gross numeraire payment at the ask when the pool sells, and the numeraire paid out when it buys. Because $Q\geq0$, external liquidation gives
\begin{equation}\label{eq:LPwealth}
V^{\rm LP}=Y+Y^f+S^bQ.
\end{equation}

\subsection{Fees retained in reserves: Uniswap v2 type}\label{sec:retained}

This subsection extends the retained-fee geometric-mean model of~\cite{FMWG3M} in two directions: the external market has proportional transaction costs, and the pool may use a general utility-indifference trading function. Geometric-mean market makers are a special case; the price-response conditions below specify the assumptions used to construct reserve dynamics for the broader class.

\subsubsection{Trading function and execution prices}

In this mechanism, the full incoming payment remains in the active reserves. Let $u$ be twice continuously differentiable and strictly concave on an open convex reserve domain $D\subset(0,\infty)^2$, with $u_x,u_y>0$. A swap with increments of opposite signs is priced by
\begin{equation}\label{eq:retained-utility-swap}
u\bigl(X+(1-\tau\1_{\{\Delta X>0\}})\Delta X,
Y+(1-\tau\1_{\{\Delta Y>0\}})\Delta Y\bigr)=u(X,Y).
\end{equation}
Both the fee-adjusted and actual post-trade reserves must lie in $D$. The swap is priced on the current utility level, but the new active reserves are $(X+\Delta X,Y+\Delta Y)$. Since $u$ is increasing, retaining the incoming fee weakly raises the utility level.

To derive the marginal prices, first let the trader sell $z>0$ units of the risky asset for a numeraire payout $v(z)$. The swap rule reads
\[
u(X+(1-\tau)z,Y-v(z))=u(X,Y).
\]
Differentiating at $z=0$, with $v(0)=0$, gives
$(1-\tau)u_x-u_yv'(0)=0$. Thus the bid is
$B=\lim_{z\downarrow0}v(z)/z=(1-\tau)u_x/u_y$.
For a purchase of $z$ units at numeraire cost $w(z)$, the rule instead reads
\[
u(X-z,Y+(1-\tau)w(z))=u(X,Y).
\]
Its derivative at zero is $-u_x+(1-\tau)u_yw'(0)=0$, so the ask is
$A=\lim_{z\downarrow0}w(z)/z=u_x/((1-\tau)u_y)$. In particular,
\begin{equation}\label{eq:utility-price}
S=s(X,Y):=\frac{u_x(X,Y)}{u_y(X,Y)},\qquad
A=\frac{S}{1-\tau},\qquad B=(1-\tau)S.
\end{equation}
Thus the no-arbitrage band \eqref{eq:band} remains valid.

\subsubsection{Reserve cash flows and fee accumulation}

For continuous finite-variation flow, the actual reserve increments satisfy
\begin{equation}\label{eq:retained-cash}
\dd X=\dd X^\uparrow-\dd X^\downarrow,\qquad
\dd Y=A\dd X^\downarrow-B\dd X^\uparrow.
\end{equation}
The total risky inventory is now $Q=X$. Applying the finite-variation chain rule to $H=u(X,Y)$ gives
\begin{equation}\label{eq:utility-growth}
\dd H=\tau u_x\dd X^\uparrow
+\frac{\tau}{1-\tau}u_x\dd X^\downarrow\geq0.
\end{equation}
Thus the current reserve curve is $u(x,y)=H_t$: the trading function is fixed and its level increases as fees accrue.

\subsubsection{Price response and reflected dynamics}
For separate fees, the monotonicity of $S=-f'(X)$ ensures that buying the risky asset lowers the pool price and selling it raises the price. Retained fees move the reserves between utility levels, so concavity along a level set alone does not ensure this response. From \eqref{eq:retained-cash}, the changes in log price per unit bought and sold by the pool are $(s_x-Bs_y)/s$ and $(-s_x+As_y)/s$, respectively. We impose
\begin{equation}\label{eq:utility-response}
a_+:=-\frac{s_x-Bs_y}{s}>0,\qquad
a_-:=\frac{-s_x+As_y}{s}>0,
\end{equation}
where $A=s/(1-\tau)$ and $B=(1-\tau)s$ are functions of the reserves. For example, $s_x<0$ and $s_y\geq0$ imply both inequalities. Under these conditions, both directions of trade restore the no-arbitrage band.

\begin{proposition}[Local retained-fee dynamics]\label{prop:utility-reflection}
Suppose $u\in C^3(D)$, \eqref{eq:utility-response} holds on $D$, $\tau+\lambda>0$, and the initial reserves belong to $D$ and lie in the no-arbitrage band. For every positive continuous external price, there is a unique continuous pure-arbitrage reserve path, of finite variation on compact time intervals before its maximal lifetime in $D$.
\end{proposition}
\begin{proof}
Write $h=-\log\gamma>0$ and $S_0=s(X_0,Y_0)$. Solve the scalar Skorokhod problem on a bounded interval, as in~\cite{KrukEtAl},
\[
Z_t=\log(S_0/S_t^*)+L_t-U_t\in[-h,h],
\]
where $L,U$ start at zero and increase only at $Z=-h$ and $Z=h$, respectively. With these continuous nondecreasing regulators fixed, solve
\begin{equation}\label{eq:utility-controlled-ode}
\dd\begin{pmatrix}X\\Y\end{pmatrix}
=\frac{1}{a_+}\begin{pmatrix}1\\-B\end{pmatrix}\dd U
+\frac{1}{a_-}\begin{pmatrix}-1\\A\end{pmatrix}\dd L.
\end{equation}
The vector fields are locally Lipschitz, since $u\in C^3$ and $a_\pm>0$. The ordinary differential equation driven by the continuous finite-variation path $(U,L)$ therefore has a unique local solution. The chain rule yields
\[
\dd\log s(X,Y)=-\dd U+\dd L,
\]
so $\log(s(X_t,Y_t)/S_t^*)=Z_t$. The mutually singular regulator measures give
\[
\dd X^\uparrow=a_+^{-1}\dd U,\qquad
\dd X^\downarrow=a_-^{-1}\dd L.
\]
Their boundary supports imply \eqref{eq:support}, and \eqref{eq:utility-controlled-ode} gives \eqref{eq:retained-cash}. Conversely, any such reserve path produces the scalar regulators by $\dd U=a_+\dd X^\uparrow$ and $\dd L=a_-\dd X^\downarrow$; uniqueness of reflection and of the differential equation gives uniqueness of the reserve path. The construction continues while the reserves remain in compact subsets of $D$. Global existence requires control of reserve boundaries and possible explosion.
\end{proof}

\subsubsection{Example: a geometric-mean market maker}
Take $u(x,y)=k\log x+\log y$ with $k>0$. Exponentiating \eqref{eq:retained-utility-swap} gives the geometric-mean rule studied in~\cite{FMWG3M}:
\begin{equation}\label{eq:retained-swap}
\bigl(X+(1-\tau\1_{\{\Delta X>0\}})\Delta X\bigr)^k
\bigl(Y+(1-\tau\1_{\{\Delta Y>0\}})\Delta Y\bigr)=X^kY.
\end{equation}
This is equivalent to using $x^\alpha y^{1-\alpha}$ with $\alpha=k/(1+k)$. The constant-product case $k=1$ is the Uniswap v2 specification.

For example, starting from $(x,y)$ with $C=x^ky$, a deposit of $z>0$ units of the risky asset leaves the numeraire reserve at
\[
y'=C\bigl(x+(1-\tau)z\bigr)^{-k}.
\]
The trader receives $y-y'$, exactly as in the separate-fee example of Section~\ref{sec:separate}. Here the full deposit remains in the active risky reserve, which becomes $x+z$. Hence
\[
C_{\rm new}=C\left(\frac{x+z}{x+(1-\tau)z}\right)^k>C
\]
when $\tau>0$. The internal price is $S=kY/X$. The current reserve curve belongs to the same family as \eqref{eq:gmm-fixed-curve}, but its scale $C=X^kY$ evolves. Equation~\eqref{eq:utility-growth} becomes
\begin{equation}\label{eq:scale-growth}
\dd\log C=k\tau\frac{\dd X^\uparrow}{X}
+\frac{k\tau}{1-\tau}\frac{\dd X^\downarrow}{X}\geq0.
\end{equation}
Here $s_x=-s/x<0$ and $s_y=k/x>0$, so \eqref{eq:utility-response} holds. For $\tau+\lambda>0$ and initial reserves in the no-arbitrage band, the reflected dynamics are explicit and global. Using the regulators from Proposition~\ref{prop:utility-reflection}, put
\begin{align*}
c_+&=1+(1-\tau)k,\qquad c_-=1+\frac{k}{1-\tau},\\
X_t&=X_0\exp\{U_t/c_+-L_t/c_-\},\\
S_t&=S_0\exp\{L_t-U_t\},\qquad Y_t=S_tX_t/k.
\end{align*}
These positive processes have finite variation on compact time intervals and satisfy \eqref{eq:retained-cash} and \eqref{eq:support}. This includes $\tau=0$, $\lambda>0$.

\section{The transaction-cost rebalancing benchmark}

\subsection{Common formulation in total holdings}\label{sec:common}

Let $Q$ and $C^{\rm LP}$ denote the liquidity provider's total risky-asset and numeraire holdings. The two mechanisms are summarized by
\[
(Q,C^{\rm LP})=
\begin{cases}
(X+X^f,Y+Y^f),&\text{separate fees},\\
(X,Y),&\text{retained fees}.
\end{cases}
\]
In both cases, writing $Q=Q_0+Q^\uparrow-Q^\downarrow$ for its Jordan decomposition, the cash equation is
\begin{equation}\label{eq:common-cash}
\dd C^{\rm LP}=A\dd Q^\downarrow-B\dd Q^\uparrow.
\end{equation}
For separate fees this follows from \eqref{eq:Q}--\eqref{eq:cash}, since $B\dd Q^\uparrow=S\dd X^\uparrow$; for retained fees it is \eqref{eq:retained-cash}. Pure arbitrage means
\begin{equation}\label{eq:common-support}
(A-S^b)\dd Q^\downarrow=0,\qquad
(S^a-B)\dd Q^\uparrow=0.
\end{equation}
Since $Q\geq0$, liquidation wealth is
\begin{equation}\label{eq:common-wealth}
V^{\rm LP}=C^{\rm LP}+S^bQ.
\end{equation}
The mechanisms generally produce different paths of total holdings from the same initial portfolio. Their common cash identity does not assert equality between those two paths.

\subsection{The benchmark and zero LVR}

For either mechanism, consider an external self-financing strategy holding the same total inventory $Q$. Its numeraire account and liquidation wealth satisfy
\begin{align}
\dd C^R&=-S^a\dd Q^\uparrow+S^b\dd Q^\downarrow,\label{eq:rebalancecash}\\
V^R&=C^R+S^bQ,\qquad C^R_0=C^{\rm LP}_0.\label{eq:rebalancewealth}
\end{align}
\begin{definition}
The loss-versus-rebalancing under external transaction costs is
\[
\operatorname{LVR}^{\rm tc}_t:=V^R_t-V^{\rm LP}_t
=C^R_t-C^{\rm LP}_t.
\]
\end{definition}
This comparison uses total inventory, including fees wherever they are held, and imposes external bid and ask execution on the benchmark. When $S^a=S^b=S^*$ it reduces to the frictionless rebalancing comparison.

\begin{theorem}[Zero LVR for both fee mechanisms]\label{thm:zero}
Suppose the external bid and ask are continuous and the total holdings are continuous and of finite variation, satisfying \eqref{eq:common-cash} and \eqref{eq:common-support}. Then
\[
V^R_t=V^{\rm LP}_t,\qquad \operatorname{LVR}^{\rm tc}_t=0.
\]
In particular, the conclusion applies to both fee mechanisms under continuous pure arbitrage, whenever their reflected dynamics are defined and $\tau+\lambda>0$.
\end{theorem}
\begin{proof}
Subtracting \eqref{eq:common-cash} from \eqref{eq:rebalancecash} gives
\begin{equation}\label{eq:lvr-dynamics}
\dd\operatorname{LVR}^{\rm tc}
=(S^b-A)\dd Q^\downarrow+(B-S^a)\dd Q^\uparrow=0.
\end{equation}
Both terms vanish by \eqref{eq:common-support}, and the initial cash accounts agree.
\end{proof}

When the pool sells, its total cash receipts equal the external bid times the inventory sold. When it buys, its total cash expenditure equals the external ask times the inventory acquired. Whether fees are kept separately or retained in reserves affects future quotes and inventory, but not this equality with the corresponding external strategy.

With no pool fee and no external transaction cost, the band collapses to $S=S^*$ on the active price range. The resulting quadratic-variation loss is described in \cite[Theorem~2 and Remark~3]{FMWLP}. To see the calculation in our notation, suppose that $X$ and $S^*$ are continuous semimartingales and work within the active range, where $Y=f(X)$ and $S^*=-f'(X)$. With zero initial fee balances, $V^{\rm LP}=f(X)+XS^*$. It\^o's formula gives
\[
\dd Y=-S^*\dd X+\frac12 f''(X)\dd\langle X\rangle,
\]
while the stochastic product rule gives
\[
\dd(XS^*)=X\dd S^*+S^*\dd X+\dd\langle X,S^*\rangle.
\]
Here the brackets denote quadratic variation and covariation. Since $S^*=-f'(X)$,
\[
\dd\langle X,S^*\rangle=-f''(X)\dd\langle X\rangle.
\]
Adding the two differentials therefore yields
\begin{equation}\label{eq:ito-lvr}
\dd V^{\rm LP}-X\dd S^*
=-\frac12 f''(X)\dd\langle X\rangle.
\end{equation}
The frictionless rebalancing wealth satisfies $\dd V^R=X\dd S^*$, so equal initial wealth implies
\[
V^R_t-V^{\rm LP}_t
=\frac12\int_0^t f''(X_u)\dd\langle X\rangle_u\geq0.
\]
The loss is strictly positive whenever this integral is positive. For concentrated liquidity, the treatment of the range endpoints uses Tanaka's formula; see \cite[proof of Theorem~2]{FMWLP}.

With either friction positive, the reflected inventory instead has continuous paths of finite variation, hence zero quadratic variation. The reserve and fee cash flows then obey the ordinary finite-variation chain rule, and the boundary identities give zero transaction-cost LVR. A positive external spread alone suffices for this change in the dynamics. The point $(\tau,\lambda)=(0,0)$ is therefore singular.

\section{Hedged wealth}

In~\cite{FMWLP,FMWG3M}, impermanent loss admits a model-free super-hedge in the frictionless external market, which is exact under continuous pure arbitrage and zero LVR. Relative to holding the initial portfolio, the hedging strategy holds $Q_0-Q_t$ risky units and starts with zero wealth. Its rebalancing trades are the same as those of the full inventory hedge holding $-Q_t$ considered below. With a positive external spread, these trades incur a cost whenever inventory changes. Thus zero transaction-cost LVR does not imply that the frictionless hedge still perfectly offsets impermanent loss. The following proposition identifies the cost as the external spread integrated against inventory turnover.

Hold $-Q$ units of the risky asset externally, with hedge cash account
\begin{equation}\label{eq:hedgecash}
\dd C^H=S^b\dd Q^\uparrow-S^a\dd Q^\downarrow.
\end{equation}
Netting the opposite risky positions before liquidation gives
\begin{equation}\label{eq:hedgedwealth}
W=C^{\rm LP}+C^H.
\end{equation}
\begin{proposition}[Cost of the external hedge]\label{prop:hedge}
Under the assumptions of Theorem~\ref{thm:zero}, for either fee mechanism,
\begin{equation}\label{eq:hedged-dynamics}
\dd W=-(S^a-S^b)(\dd Q^\uparrow+\dd Q^\downarrow)
=-(S^a-S^b)\dd\operatorname{TV}(Q).
\end{equation}
Here $\operatorname{TV}(Q)$ denotes the total variation of $Q$. Thus $W$ is nonincreasing.
\end{proposition}
\begin{proof}
Adding \eqref{eq:common-cash} and \eqref{eq:hedgecash} yields
\[
\dd W=(A-S^a)\dd Q^\downarrow+(S^b-B)\dd Q^\uparrow.
\]
Apply \eqref{eq:common-support}.
\end{proof}

\section{Liquidity provision as an investment strategy}\label{sec:representation}

Theorem~\ref{thm:zero} identifies a pool's liquidation wealth with that of the external strategy holding its total risky inventory. To implement a desired strategy, it therefore remains to choose a pricing rule that generates the required holdings. We use retained fees so that the actual reserves $(X,Y)$ are the strategy portfolio. We first prescribe an inventory schedule $X=q(S)$ and then consider constant no-trade boundaries for portfolio weights.

\subsection{Choosing an inventory schedule}

Let $q$ specify the desired risky inventory at each internal price. The reserve domain will be its image, so both the inventory range and the associated curve are determined by the investor's choice of $q$.

\begin{proposition}[Representation of inventory rules]\label{prop:representation}
Choose a continuously differentiable inventory schedule $q:(0,\infty)\to(0,\infty)$ with $q'<0$, and assume that
\begin{equation}\label{eq:finite-funding}
\int_0^{p_0}q(r)\dd r<\infty
\end{equation}
for some $p_0>0$. Let $I=q((0,\infty))$ be its inventory range. For any numeraire buffer $\eta\geq0$, the formula
\begin{equation}\label{eq:q-designed-curve}
f(q(p))=\eta+\int_0^p q(r)\dd r-pq(p),\qquad p>0,
\end{equation}
defines a twice continuously differentiable, strictly convex decreasing function $f:I\to[0,\infty)$ satisfying
\[
q(p)=(-f')^{-1}(p),\qquad p>0.
\]
Moreover, $y=f(x)$ is a level set of a strictly concave function which is increasing in both arguments. Thus $q$ determines an admissible inventory--price rule, with $I$ determined by the choice of $q$.
\end{proposition}

\begin{proof}
Since $q$ is continuous away from zero, \eqref{eq:finite-funding} implies that the integral in \eqref{eq:q-designed-curve} is finite for every $p>0$. Strict decrease makes $q$ a bijection onto the interval $I$. Hence \eqref{eq:q-designed-curve} defines $f$ uniquely for the chosen buffer $\eta$. For $0<r<p$, $q(r)\geq q(p)$, so
\[
\int_0^p q(r)\dd r\geq pq(p).
\]
Consequently $f(q(p))\geq\eta\geq0$. Differentiating \eqref{eq:q-designed-curve} gives
\[
f'(q(p))q'(p)=-pq'(p),
\]
and therefore
\[
f'(q(p))=-p<0,\qquad
f''(q(p))=-\frac{1}{q'(p)}>0.
\]
The inverse function theorem justifies these derivatives because $q'<0$. Thus $f$ is twice continuously differentiable, strictly decreasing and strictly convex, with $(-f')^{-1}=q$.

To recover a utility on the open convex domain $D=I\times(0,\infty)$, put
\[
u(x,y)=-\exp\{-(y-f(x))\}.
\]
It is increasing in both arguments because $f'<0$. Its Hessian is negative definite because $f''>0$, and hence $u$ is strictly concave. Strict decrease of $q$ makes the integral inequality above strict, so $f(x)>0$ on $I$ and the curve $y=f(x)$ lies in $D$. It is the level set $u=-1$.
\end{proof}

This formulation starts from the desired inventory schedule and lets its image determine the reserve domain. There is no need to prescribe $I$ in advance: $q$ may have a bounded or unbounded range. The integrability assumption limits how rapidly inventory can grow near zero price. For example, $q(p)=c p^{-\beta}$ with $c>0$ and $0<\beta<1$ is admissible and, with $\eta=0$, yields a geometric-mean curve with $k=(1-\beta)/\beta$. A bounded schedule near zero is also admissible, whereas $q(p)=1/p$ on the entire positive axis is not.

\subsection{Implementing the inventory schedule with retained fees}\label{sec:q-retained}

The utility constructed in Proposition~\ref{prop:representation} has marginal price $-f'(x)$ at every numeraire reserve $y$. Retaining fees therefore preserves the prescribed relation between total inventory and internal price.

\begin{proposition}[Retained-fee implementation of an inventory rule]\label{prop:q-retained}
Let $q$ and $f$ satisfy Proposition~\ref{prop:representation}, and on $D=I\times(0,\infty)$ set
\[
u(x,y)=-\exp\{f(x)-y\}.
\]
Use the retained-fee swap rule \eqref{eq:retained-utility-swap}. Then
\begin{equation}\label{eq:q-retained-price}
S=\frac{u_x}{u_y}=-f'(X),\qquad Q=X=q(S).
\end{equation}
The price-response condition \eqref{eq:utility-response} holds automatically, with
\[
a_+=a_-=\frac{f''(X)}{S}=-\frac{1}{S q'(S)}>0.
\]
For continuous flow, the excess numeraire reserve $K:=Y-f(X)$ satisfies
\begin{equation}\label{eq:excess-cash}
\dd K=\tau S\dd X^\uparrow
+\frac{\tau}{1-\tau}S\dd X^\downarrow\geq0.
\end{equation}
If $K_0\geq0$, the reserves remain nonnegative. For a positive continuous external price and $\tau+\lambda>0$, the pure-arbitrage dynamics exist globally from any initial state in $D$ with $K_0\geq0$ in the no-arbitrage band. Their liquidation wealth agrees with that of the external self-financing strategy holding $q(S)$ and starting from the same holdings.
\end{proposition}
\begin{proof}
Differentiating the utility gives $u_x/u_y=-f'(x)$, independent of $y$. The price rule follows from Proposition~\ref{prop:representation}. Hence $s_x=-f''(x)$ and $s_y=0$, so both coefficients in \eqref{eq:utility-response} equal $f''(x)/s(x,y)$. Proposition~\ref{prop:representation} gives $f''(q(p))=-1/q'(p)>0$, proving the displayed formula and its positivity. Subtracting $\dd f(X)=-S\dd X$ from \eqref{eq:retained-cash} gives \eqref{eq:excess-cash}. Thus the evolving level sets are vertical translates $y=f(x)+K_t$, and $Y_t\geq f(X_t)\geq0$ when $K_0\geq0$.

A direct construction avoids the additional $C^3$ assumption in Proposition~\ref{prop:utility-reflection}. Reflect $\log(S_0/S_t^*)$ on $[-h,h]$ as above and set
\[
S_t=S_0e^{L_t-U_t},\qquad X_t=q(S_t).
\]
Since $q'<0$, the Jordan increments are
\[
\dd X^\uparrow=-q'(S)S\dd U,\qquad
\dd X^\downarrow=-q'(S)S\dd L.
\]
Define $K$ by \eqref{eq:excess-cash} and $Y=f(X)+K$. These processes have finite variation on every compact time interval: $S$ stays in a compact subset of $(0,\infty)$ there, on which $q$ and $q'$ are bounded. The construction satisfies \eqref{eq:retained-cash} and the boundary conditions \eqref{eq:support}. Strict decrease of $q$ makes $f(q(p))>0$ for every $p>0$, even with zero buffer. Thus the reserves stay in the interior on compact time intervals, with no finite-time boundary exit or explosion, and Theorem~\ref{thm:zero} yields the wealth equality.
\end{proof}

The quantity $K$ is held within the actual numeraire reserve. Since the marginal price is independent of $K$, fee accumulation preserves $X=q(S)$. The actual reserve value at internal price $p$ is $f(q(p))+K+pq(p)$, including this accumulated excess numeraire.

For a power-law schedule, this construction and a geometric-mean AMM can share the same initial reserve curve yet have different subsequent dynamics. Here the curve shifts vertically and $X=q(S)$ stays fixed as a rule. In the geometric-mean model its scale grows, keeping a constant risky fraction at the internal price. The former is suited to a prescribed absolute inventory schedule; the latter, used next, is suited to prescribed portfolio-weight boundaries.

\subsection{Black--Scholes optimal investment}

For portfolio-weight rules, a geometric-mean pool provides a different design. Work in discounted units and suppose that the external reference price follows
\[
\frac{\dd S_t^*}{S_t^*}=\mu\dd t+\sigma\dd\mathcal{B}_t.
\]
Here $\mathcal{B}$ is a standard Brownian motion and $\sigma>0$. Consider an infinite-horizon pure investment problem with constant relative risk aversion in a parameter regime where the optimal policy has constant interior no-trade boundaries
\[
0<\pi_-<\pi_+<1.
\]
The investor does not trade while the risky fraction lies in $(\pi_-,\pi_+)$, buys at $\pi_-$, and sells at $\pi_+$. We define the risky fraction using reference-price wealth, $\pi=XS^*/(Y+XS^*)$. The boundaries must be obtained from the free-boundary problem for the chosen objective and this price convention. Davis and Norman~\cite{DavisNorman} and Shreve and Soner~\cite{ShreveSoner} provide classical investment--consumption formulations. Here we take constant boundaries $(\pi_-,\pi_+)$ as given and implement their self-financing trading rule; a consumption policy would additionally require withdrawals.

Put
\[
a=\frac{\pi_-}{1-\pi_-},\qquad
b=\frac{\pi_+}{1-\pi_+},
\qquad
k=\sqrt{ab},\qquad
\gamma=\sqrt{\frac{a}{b}}.
\]
Consider the reserve curve
\begin{equation}\label{eq:merton-curve}
f_t(x)=C_t x^{-k},\qquad C_t=X_t^kY_t>0.
\end{equation}
Use the retained-fee mechanism of Section~\ref{sec:retained}. The actual reserves are the entire strategy holdings, and $C_t$ evolves by \eqref{eq:scale-growth}. The internal price and internal-price risky fraction are
\[
S_t=k\frac{Y_t}{X_t},\qquad
\alpha=\frac{X_tS_t}{Y_t+X_tS_t}=\frac{k}{1+k}.
\]
At an external reference price $p$, the risky fraction of reserves is
\[
\pi_t(p)=\frac{X_tp}{Y_t+X_tp}.
\]
The no-arbitrage band \eqref{eq:band} is equivalent to
\[
\gamma\leq\frac{p}{S_t}\leq\frac1\gamma.
\]
Therefore the risky fractions at its lower and upper boundaries are
\[
\frac{k\gamma}{1+k\gamma}=\pi_-,
\qquad
\frac{k}{k+\gamma}=\pi_+.
\]
The pool leaves its holdings unchanged in the interior, buys the risky asset at the lower weight boundary, and sells it at the upper weight boundary.

To obtain the required band, choose the pool fee from
\begin{equation}\label{eq:optimal-fee}
(1-\tau)(1-\lambda)=\gamma,
\qquad\text{that is}\qquad
\tau=1-\frac{\gamma}{1-\lambda}.
\end{equation}
This is feasible when $\gamma\leq1-\lambda$. The external spread itself imposes a minimum band; a narrower target cannot be implemented without a rebate. The initial value $C_0$ fixes the initial curve level, while $k$, or equivalently $\alpha$, determines the center of the optimal wedge. If the numeraire earns the short rate, it may be represented by an interest-bearing token; equivalently, all quantities may be discounted as above.

At the buying boundary, \eqref{eq:retained-cash} becomes $\dd Y=-S^a\dd X^\uparrow$; at the selling boundary it becomes $\dd Y=S^b\dd X^\downarrow$. The pool therefore implements the self-financing reflected trading rule for the prescribed weight interval, starting from the same feasible holdings. Theorem~\ref{thm:zero} identifies its liquidation wealth with that external strategy. This is an implementation statement conditional on the target boundaries being optimal for the investor's specified problem and satisfying \eqref{eq:optimal-fee}.

The constant-product restriction of Uniswap v2 fixes $k=1$; general target boundaries require a geometric-mean pool with the chosen weight and fee. Thus ``Uniswap v2 type'' refers here to retaining fees, not to an assertion that an unmodified Uniswap v2 pool can implement every target.

For separate fees, the bare reserve ratio still has the displayed boundary fractions, but the investor's total fraction is
\[
\frac{(X+X^f)p}{Y+Y^f+(X+X^f)p}.
\]
It generally differs from the active-reserve fraction. Exact implementation with separate accounts would require continuous management reproducing the retained-fee dynamics, including the evolving scale.

\paragraph{Scope of the constructions.}
Both designs use a fixed trading function. Strategies depending on time, past prices, or additional state variables may require an adapted pricing rule. Stateful hooks, such as those provided by Uniswap v4, offer a possible implementation route through custom accounting and pricing~\cite{UniswapV4}. Establishing an exact strategy implementation in that setting requires a separate analysis. The results here concern continuous reserve dynamics and do not include gas costs, contract risk, or liquidity deposits and withdrawals.

\section{Beyond pure continuous arbitrage}

Theorem~\ref{thm:zero} isolates the continuous arbitrage mechanism. Two extensions change the conclusion.

First, liquidity-taker flow need not occur on the boundaries \eqref{eq:support}. For either fee mechanism with continuous finite-variation holdings satisfying \eqref{eq:common-cash},
\begin{equation}\label{eq:general-lvr}
\dd\operatorname{LVR}^{\rm tc}
=(S^b-A)\dd Q^\downarrow
+(B-S^a)\dd Q^\uparrow\leq0
\end{equation}
under the no-arbitrage inequalities \eqref{eq:na}. Thus order flow in the interior makes the liquidity provider outperform the external rebalancing benchmark. Equality holds for the arbitrage component because it is supported on the boundary. For separate fees this recovers the reserve-coordinate formula using \eqref{eq:Q}; for retained fees, $Q=X$.

Second, a jump in the external price may move the pool outside the no-arbitrage band. Arbitrageurs then profit by trading against the pool's stale quotes and taking the opposite position in the external market. Relative to retaining the pre-trade portfolio, valued at the new external reference price, the liquidity provider suffers a loss from these unfavorable exchanges, even after collecting pool fees. This loss funds the arbitrage profit and the external execution costs. The trade traverses a nonzero range of pool prices, so the boundary execution identities used in the continuous-time proof no longer hold throughout the trade. The zero-LVR conclusion therefore does not extend directly to price jumps.

\section{Conclusion}

External transaction costs change both the no-arbitrage region and the appropriate rebalancing benchmark. Under continuous external prices and pure arbitrage, a pool with either separate or retained fees has the same liquidation wealth as the external self-financing strategy holding its total risky inventory whenever the pool fee or the external spread is positive. Zero LVR does not imply costless hedging: the external hedge loses the spread integrated against turnover for both positions.

Retained fees also permit direct investment implementation. Utility-indifference pools with vertically shifting reserve curves preserve prescribed inventory schedules, while geometric-mean pools implement feasible constant no-trade boundaries for portfolio weights. These two designs connect the flexibility available to liquidity providers with explicit trading rules enforced by arbitrage.

\paragraph{Acknowledgement.}
This work was supported by JSPS KAKENHI Grant Number 25K07112.

\paragraph{Declaration on the use of generative AI.}
The mathematical ideas and arguments were developed under the author's direction, and the author checked all mathematical statements and proofs. OpenAI's GPT-6, accessed through Codex, assisted with drafting and refining the exposition of these arguments, language editing, reference checks, and LaTeX preparation. It also assisted with supplementary consistency checks and partial Lean formalization of selected results. These checks do not constitute a full formal verification of the manuscript. The author takes responsibility for the final content.

\end{document}